**Reducing Richtmyer Meshkov Instability Jet Velocity via Inverse Design**

Dylan J. Kline[1a*], Michael P. Hennessey[1a], David K. Amondson[1], Steve Lin[1], Michael D. Grapes[1], Massimiliano Ferrucci[1], Peggy Li[1], H. Keo Springer[1], Robert V. Reeves[1], Kyle T. Sullivan[1], Jonathan L. Belof[1]

[1]Lawrence Livermore National Laboratory, Livermore, CA 94550, USA
[a]These authors contributed equally to this work.
* Corresponding author: kline11@llnl.gov

**ABSTRACT**

In this work, we detail a novel application of inverse design and advanced manufacturing to rapidly develop and experimentally validate modifications to a shaped charge jet analogue. The shaped charge jet analogue comprises a conical copper liner, high explosive (HE), and silicone buffer. We apply a genetic algorithm to determine an optimal buffer design that can be placed between the liner and the HE that results in the largest possible change in jet velocity. The use of a genetic algorithm allows for discoveries of unintuitive, complex, yet optimal buffer designs. Experiments using the optimal design verified the effectiveness of the buffer and validated the modeling.

## 1. INTRODUCTION

Controlling physical instabilities is a grand challenge in science and engineering. Fluid instabilities, and more specifically interfacial instabilities, are among the most widely studied physical phenomena as they regularly manifest when mixing fluids with different densities. A classic example of interfacial instabilities, the Rayleigh-Taylor instability (RTI), is used to describe the evolution of an interface between fluids of different densities as the denser fluid is accelerated into the lighter one [1–3]. In an RT unstable interface, the initial growth of the interfacial perturbation amplitude is exponential and exhibits "spikes" in the interface before turbulent mixing [1–3]. However, in the case where the materials are impulsively accelerated into each other (typically under shock conditions), the interfacial evolution fundamentally changes. The Richtmyer-Meshkov instability (RMI) is the impulse-accelerated case of the RT instability [3–6]. Like RTI, RMI is characterized by the evolution of spikes in the interface, however in RMI, the evolution of the instability is not dependent on the original shock direction (i.e. not dependent on light-to-heavy or heavy-to-light fluid shock propagation) and the initial growth of the spikes is linear [3].

RMI has been of particular interest for its role in materials undergoing extreme dynamic loading. RMI is prominent across multiple disciplines, especially including astrophysics [3,7], combustion [8], and shaped charges [9–12]. More recently, RMI has been investigated for its role in inertial confinement fusion [13–15]. The evolution of the instability can lead to unpredictable behavior in the materials being mixed and can yield undesirable side effects (e.g. inefficient reactions during fusion), thus it is critical to develop methods to predict and control the instability.

A promising route to address the ongoing challenge in controlling instabilities leverages advancements in computing capabilities and manufacturing methods. From the computational perspective of this problem, machine learning is well-suited for rapidly exploring the feasible design space for designing geometries that could yield tailored instability evolution whether it be for mitigating or augmenting the spikes [11,12,16,17]. However, since such an optimization procedure can effectively explore an "unlimited" design space within the problem, one could expect that the optimal design for an objective result may be difficult (if not impossible) to manufacture. Considering the possible complexity, additive manufacturing (AM) is an attractive method to manufacture components with radical designs [18–20]. By coupling inverse design and additive manufacturing, it may be possible to engineer dynamic material response that can be tailored to any application.

In this article, we demonstrate an example of applying a computational optimization technique to design a part with a tailored RMI evolution, the fabrication of the parts with AM techniques, and the performance of the parts compared to simulation results. We focus specifically on a method to mitigate the RMI evolution in an explosively-driven linear shaped charge analogue using a computationally-optimized buffer. We begin by describing our sample problem and the computational methods used to design the buffer. We then discuss the fabrication of the parts using AM techniques and demonstrate the effectiveness of the buffer in high explosives (HE) testing. Our approach demonstrates how next-generation design and manufacturing techniques can be coupled to solve grand challenges in physics.

## 2. METHODS

### 2.1. HIGH EXPLOSIVES DETONATION EXPERIMENT

#### 2.1.1. EXPERIMENT DESIGN

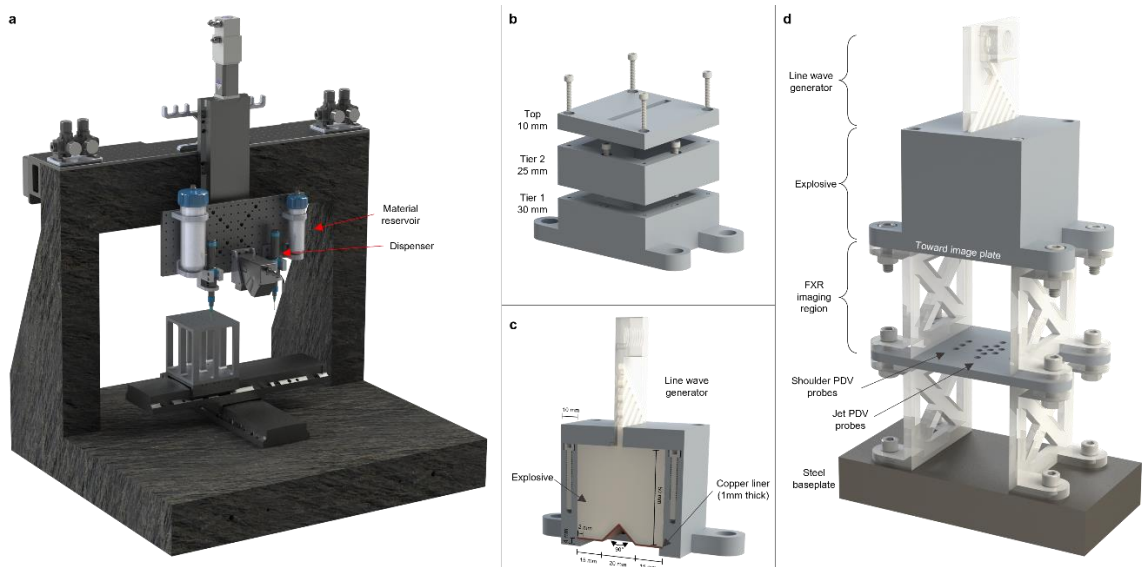

Figure 1: Manufacturing method and experiment design details. (a) Custom 3D printer used for component fabrication. (b) Assembly instructions for external casing of the experiment. After filling the first tier, the second tier is added and filled. Top is added after filling and curing parts. Tier 1 has a copper liner placed in before printing. (c) Cross sectional image of the experiment and dimensions. Configuration shown is for the baseline experiment. (d) Final assembly configuration for detonation experiment (steel shrapnel protection not shown).

Prior to simulating or fabricating any components, multiple key considerations were made regarding the experimental design. The goal of the experiment was to verify that a computationally-optimized silicone buffer can deform a shock front to reduce a shaped charge's jet velocity without substantially reducing a reference velocity. Shock wave deformation and early deformation of the copper liner would be imaged using flash X-ray (FXR) radiography. To achieve the best image quality in the radiographs, the experiment was designed as a linear shaped charge. This design would enhance the contrast in the image between the jetting material and free space, although it would result in a parallax effect for any off-center images. Photonic doppler velocimetry (PDV) was also to be used for higher-resolution velocity data of jet and reference velocity regions. The experiment would be initiated using a single detonator placed against a line wave generator (LWG) spanning the entire width of the explosives in the assembly to more closely match 2D simulations. To minimize potential damage to the experiment enclosure, the total explosive mass was limited to

250 g and the experiment would be placed atop six 1" thick steel plate spaced 1" apart. This constrained the experiment's explosive volume to ~140 cm$^3$. Components were sized to have a volume ≤ 120 cm$^3$ for engineering safety and design simplicity. The thickness of the liner was chosen to be 1 mm and would feature a 90° liner angle with an output region spanning 20 mm. After incorporating the liner to the experimental design, the total volume of the HE would be 120 cm$^3$ (mass ~210 g). Early simulations suggested that the peak velocity of the shoulder in the control experiment would be reached within 50 mm of its original position, therefore we placed the PDV array ~60 mm away from the reference velocity region of the liner. The entire assembly was also mounted to a 1" thick steel baseplate which would act as a weight to keep the center of gravity low and enhance the shrapnel protection. A rendering of the final experimental design can be seen in Figure 1. Detailed discussions of the fabrication methods and diagnostics are provided in the following sections.

### 2.1.2. MATERIALS

Two materials were used in the internal cavity of the additively manufactured shaped charge: (1) an extrudable high-power explosive (Lawrence Livermore National Laboratory) and (2) an inert silicone, DOWSIL SE1700 (Dow, Inc.). The silicone is easily extruded through nozzles and is chemically compatible with the HE [21]. The silicone was dyed blue using SilcPig pigments (Smooth-On) to visually distinguish the material from the HE during printing operations. Both materials are mixed and degassed in large cups using a planetary centrifugal mixer (Flacktek DAC 600.2 SpeedMixer) prior to being loaded on the printer to remove any trapped air. After the initial mix stage, materials are transferred to fluid cartridges (Nordson EFD) and are spun for a second time in the mixer to remove air pockets. A piston is added to each cartridge prior to being loaded into a retainer system, mounted to the 3D printer, and pressurized. Custom 1 mm thick oxygen-free, high conductivity copper liners were procured for this experiment and were used as-received from the manufacturer (Protolabs). The assembly is initiated using an RP-2 exploding bridge wire detonator (Teledyne Defense Electronics) and a LWG filled with a PETN-based explosive (Lawrence Livermore National Laboratory).

### 2.1.3. PART FABRICATION

Two components were prepared for this experiment to evaluate the effectiveness of the computationally optimized buffering technique. The first component served as the control experiment and contains a monolithic HE charge. The second component contains the same HE but featured a silicone buffer placed between the HE and the copper liner (design techniques described below). Both components were made via a direct ink write (DIW) process on a custom 3D printer (see Figure 1a). The 3D printer uses 2 Aerotech PRO165LM for the X and Y travel and an Aerotech PRO165SL for Z travel. The axes are driven by Aerotech DP3200E servo drives housed in Aerotech Npaq drive rack which is controlled using the Aerotech A3200 software motion controller. Material dispensing and axis movement are controlled using Aerotech AeroBasic commands [22]. Material reservoirs and dispensers are mounted to a breadboard on the Z axis using custom fixtures. Both materials are pressure-fed to separate progressive cavity pumps (Viscotec preeflow eco-PEN450) and out through straight-barrel Luer lock nozzles (Nordson EFD).

Material properties and dispensing equipment dimensions primarily drove the fabrication process. The explosive is space filling and does not retain its structure after being extruded, thus the components were printed into 3D printed plastic molds (Stratasys). To ensure good contact between the printed materials and copper liner, materials were printed directly onto the liner which was glued into the mold. The print program was also split into multiple 25 mm tall segments to overcome the limited nozzle length (25.4 mm); after the first tier of the print was filled, a second tier was added, and the print routine was resumed.

Component toolpaths were prepared using commercial slicing software (Simplify3D) with an emphasis on creating parts with high density infills. The layer height and nominal filament width of the silicone was 0.5 mm and 1.27 mm, respectively. The layer height and nominal filament width of the explosive was 1.0 mm and 1.83 mm, respectively. The layer height of the silicone was chosen to be smaller to improve the printed resolution of the buffer and to form "wells" in the part which could then be filled by the space-filling explosive. Toolpaths were then post-processed for compatibility with the 3D printer. Prior to printing, the offset between the two nozzles is measured and a extrusion timing calibration routine is run to minimize the defects in the part [21,23]. After the print is completed, the parts were thermally cured in an oven at 50°C for 3 days.

Once the parts were cured, the LWG and detonator are added to the assembly. The explosive in the LWG is hand-loaded and fits into the top layer of the box such that the output is centered along centerline of the part parallel to the jetting region of the copper liner. Prior to installing the LWG and the top of the case, the top surface of the 3D printed component is manually trimmed flat.

### 2.1.4. DIAGNOSTICS

Key diagnostics for the detonation experiments performed were flash X-ray radiography (FXR) and photonic doppler velocimetry (PDV). FXR is necessary for resolving detonation fronts and material deformation in a 2D image for direct comparison to simulations. PDV can be used to extract point velocimetry data at a higher rate and resolution than can be achieved with FXR.

The FXR system used three 450 kV pulsers (L3 Harris) which were positioned radially on a single port of the 10 kg spherical tank in LLNL's High Explosives Applications Facility (8° spacing). The X-rays pass through a steel collimator mounted on the interior of the tank and then through a second collimator mounted to a cylindrical shrapnel catcher. After passing through the experiment and the encasing shrapnel catcher, X-rays are captured on a single digital imaging plate housed in a protective case. An example layout of the tank setup can be seen in Figure 7. FXR trigger timing was predetermined to capture the phenomena of interest based on hydrodynamic simulations. Triggering was controlled via an in-house program. Since the experiment performed is not radially symmetric, radiographs captured in the +/- 8° positions exhibited a parallax effect. We will discuss methods for creating simulated radiographs in a later section. Image analysis of the parts was done with ImageJ [24].

Velocimetry data was captured using an array of 15 collimator PDV probes (OzOptics LPC-01-1550-9/125-S-0.95-5AS-60-3A-3-5) placed ~60 mm from the flat region of the copper liner (spot size 0.95 mm). As can be seen in Figure 1d, 5 probes were directed along the center of the copper liner, 2 probes were placed on either side of the center at the halfway position between the center and reference velocity region, and 3 probes were placed on each flat surface. A 1550 nm laser (IPG Photonics) is passed through the probes and the return signal is captured using a GHz oscilloscope (Tektronix DPO73304DX). Velocimetry data was extracted using a custom program (LLNL Advanced PDV OpsV3) and postprocessed with Python routines.

### 2.1.5. PART INSPECTION

X-ray computed tomography (XRCT) was performed on the parts prior to final assembly. The component was placed atop a rotating stage and imaged using a 450 kV source (Comet Yxlon) and a digital detector (Perkin Elmer XRD 1620). A 3D reconstruction of the part was prepared using Livermore Tomography Tools [25].

### 2.2. HYDRODYNAMIC SIMULATIONS AND COMPUTATIONAL METHODS

### 2.2.1. SIMULATION SETUP

To understand the evolution of the detonation within this device and to optimize the design for a jet-mitigating buffer, we simulate the device using the LLNL hydrocode ALE3D [26]. As a first pass, a 2D slice of the device was taken and discretized on a grid at a resolution of 100 zones/cm (Figure 2). The LWG inlet is modeled as an HMX-based HE undergoing programmed burn, where the HE is modeled using a JWL equation of state. The LWG inlet initiates the main charge and its detonative wave subsequently interacts with the copper liner. The HMX-based main charge is modeled using a reactive flow model in the thermochemical equilibrium code CHEETAH [27] to evolve the detonation products species at each time-step. The copper liner is modeled using the Steinberg-Guinan strength model (Equation 1),

$$Y = Y_0 f(\epsilon_p) \frac{G(P,T)}{G_0}, \quad \text{with } Y_0 f(\epsilon_p) = Y_0 \left[1 + \beta(\epsilon_p + \epsilon_0)\right]^n \leq Y_{max}$$

$$\text{where } G(P,T) = \left[G_0 + a_P G_0 P \left(1 - a_n + a_n \left(\frac{\rho}{\rho_0}\right)^{-\frac{1}{3}}\right) - G_0 a_T (T - 300)\right], \quad (1)$$

together with a simple stress-based spall model and a Mie-Gruneisen equation of state (Equation 2),

$$p = \frac{\rho_0 c_0^2 \mu \left[1 + \left(1 - \frac{\gamma_0}{2}\right)\mu - \frac{b}{2}\mu^2\right]}{\left[1 - (S_1 - 1)\mu - S_2 \frac{\mu^2}{\mu + 1} - S_3 \frac{\mu^3}{(\mu + 1)^2}\right]^2} + (\gamma_0 + b\mu)E, \quad \text{with } \mu = \frac{\rho}{\rho_0} - 1 \quad (2)$$

and parameters in Tables 1 and 2, respectively. The spall model in use detects if the maximum principal stress exceeds a value of 1.2 GPa. If spall is detected, the deviatoric stresses are set to zero in that numerical zone. The lucite case is modeled using a Steinberg-Guinan strength model and Mie-Gruneisen equation of state with parameters in Tables 1 and 2. The buffer itself is modeled as Sylgard using a tabular equation of state [28].

Table 1: Steinberg-Guinan strength model parameters.

| Material | $Y_0$ | $\beta$ | $\epsilon_0$ | $n$ | $Y_{max}$ | $G_0$ | $a_P$ | $a_n$ | $a_T$ |
|---|---|---|---|---|---|---|---|---|---|
| Copper | $4.0E-5$ | 1200 | 0 | 0.8 | $6.4E-3$ | 0.477 | 2.83 | 1 | $3.77E-4$ |
| Lucite | $4.2E-3$ | 0 | 0 | 1 | $4.2E-3$ | 0.0232 | 0 | 0 | 0 |

Table 2: Mie-Gruneisen equation of state parameters.

| Material | $\rho_0$ | $c_0$ | $\gamma_0$ | $b$ | $S_1$ | $S_2$ | $S_3$ |
|---|---|---|---|---|---|---|---|
| Copper | 8.93 | 0.394 | 2.02 | 0.47 | 1.489 | 0 | 0 |
| Lucite | 1.182 | 0.218 | 0.85 | 0 | 2.088 | $-1.124$ | 0 |

While the 2D simulations were used to computationally optimize the design of our jet-mitigating buffer, 3D simulations were performed to better capture the physics of the experimental setup. In moving to 3D, the highest resolution we were able to utilize was 40 zones/cm and the main charge was modeled using the JWL++ reactive flow model in place of CHEETAH. The LLNL X-ray radiography code HADES was used in conjunction with the 3D simulations to generate simulated radiographs to compare to the experimental results [29].

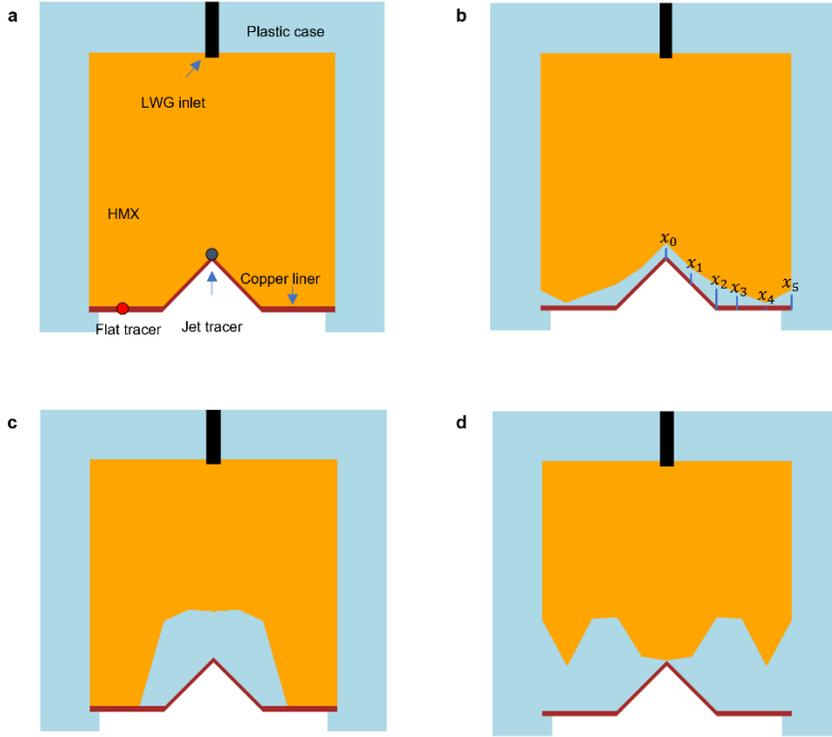

Figure 2: (a) An illustration of the 2D simulation setup. The LWG is in black, the main charge in orange, the copper liner in red and plastic case in light blue. The jet tracer and flat tracer locations are illustrated as a blue and red circle placed on the copper liner. (b) An illustration of the geometry of the inert buffer as defined by 6 points in space in light blue. (c) A naïve buffer design. (d) The computationally optimized buffer design.

### 2.2.2. COMPUTATIONAL OPTIMIZATION ALGORITHM

The optimization of the buffer insert geometry was performed using the genetic algorithm differential evolution, by modifying the x value of the 6 points in y along the rear of the liner (Figure 2b). This algorithm is an unsupervised learning parallelized stochastic direct search method which begins with an initial population of M parameter vectors $x_{i,0}$ for $i = 1,2,...,M$, randomly sampled from an initial bound using Latin Hypercube Sampling. The objective function, $f(x_{i,n})$, is evaluated for each parameter vector in parallel then the entire population is updated to new sets of parameters via mutation, crossover, and selection. While multiple options for mutation operations have been developed, we have chosen the variant known as "best/1/bin."

The mutation in this variant generates mutant parameter vectors, $v_{i,n}$, that are based on the parameters of the current best parameter set, $x_{r_b,n}$, and two other parameter sets chosen randomly such that Equation 3 is true.

$$v_{i,n} = x_{r_b,n} + F(x_{r_1,n} - x_{r_2,n}), \qquad r_b \neq r_1 \neq r_2 \qquad (3)$$

Here, $F \in [.5,1]$ is an amplification factor randomly chosen at each generation. A new trial parameter set, $u_{i,n}$, is then formed through crossover between the mutant vectors and the members of the prior parameter set. The crossover operation randomly chooses parameter values from either the mutant vector or the previous parameter set for each parameter. If the new trial vectors have an objective function value smaller than the prior parameter set, then the old parameter sets are replaced, i.e.,

$$x_{i,n+1} = \begin{cases} u_{i,n} & if \quad f(u_{i,n}) < f(x_{i,n}) \\ x_{i,n} & otherwise. \end{cases} \quad (4)$$

This process continues until the minimum objective function value throughout the population is determined to converge.

The objective function in our case is determined to be the weighted average of the jet velocity and reciprocal velocity of the flat of the copper liner as measured by tracer particles placed on the liner face as depicted in Figure 2a (given by Equation 5).

$$f(x) = 7.5 v_{jet} + \frac{1}{v_{flat}} \quad (5)$$

Thus, by minimizing the objective function, we balance the minimization of the jet velocity with the maximization of the 'bulk' velocity as measured by the velocity of the flat of the liner.

## 3. RESULTS AND DISCUSSION

### 3.1. HYDRODYNAMIC SIMULATIONS

We begin presenting our results for the baseline design which does not utilize a buffer. For each design discussed, we investigate the visual appearance of the jetting process together with a look at the simulated PDV velocities.

For the baseline case, the jetting occurs following the interaction of the detonation wave in the main charge with the copper liner (Figure 2a, simulation results in Figure 3). At early times following the interaction, a small jet-slug feature develops at the apex of the liner. The small jetting feature forms from the liner as the surface inverts. Meanwhile, the slug forms from the compressed bulk of the liner as the detonation front pushes the rays of the 'V' section of the liner towards one another. During this process, the 'V' section of the liner is also imparted with a strong vertical velocity from the detonation wave. Following the initial jet formation, we see a growth of the jet-slug feature in both directions. The slug continues to be formed along the vertical axis by the compression of the 'V' in the horizontal direction following the passage of the detonation front. The jet continues to grow pulling in material from the 'V' section of the liner. The 'V' section of the liner separates from the flat sections of the liner following the interaction of the detonation front with the corners, which we call the 'elbow break'. This allows for the jet to pull in more liner material due to vortical forces, which we call 'jet coalescence'. Eventually all the liner material that made up the 'V' section of the liner is pulled into the jet-slug feature.

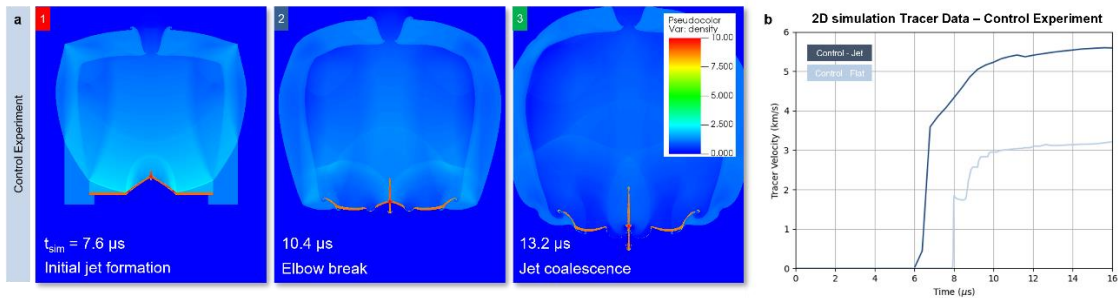

Figure 3: (a) Images 1, 2 and 3 show the time evolution of the baseline charge simulation in density and (b) the tracer velocity results.

The velocity measurements help confirm these observations. As the simulated PDV measurements are taken at the surface of the liner, we see an initial peak at the apex of the liner at about 6.5 µs following detonation and an initial peak in the flat region about 1 µs later corresponding to the detonation front travel time between the apex and flat of the liner. We see oscillations in velocity in the flat of the liner following the initial jump-off corresponding to "ringing" in the copper. Eventually, the flat reaches a steady velocity at ~11 µs following detonation.

Interestingly, rather than having a single jump-off, the jet velocity trace shows a jump-off followed by a gradual velocity increase over the following 4 µs. The initial jump-off corresponds to the impact of the detonation front and the inversion of the surface forming the initial jet, and the gradual velocity increase corresponds to the jet coalescence process discussed above. A first approach to mitigating the jet is to simply reduce the energy content of our charge by placing a piece of inert material behind the 'V' section of the liner (Figure 2c, simulation results in Figure 4). However, simulations of these components show that placing a bulk piece of silicone behind liner apex liner can increase the jet velocity depending on the shape of the inert material. The mechanism behind the jet velocity increase is the appearance of a Mach stem within the inert material. This Mach stem both deforms the wave front due to the higher impedance in the inert and produces a large region of high kinetic energy. As the impedance mismatch between the inert material and the copper is less severe than the impedance mismatch between the HE and the copper, the energy from the inert material is imparted into the copper more strongly. It is this kinetic energy transfer that drives the jetting process. Due to the shape of the inert insert, the compression of the rays of the 'V' section of the liner is driven with more energy, resulting in a jet coalescence effect that produces a faster moving jet (compared to the baseline case). The velocity trace agrees with these observations and shows that the removal of the HE behind the 'V' of the liner does result in less bulk motion, according to the lower velocity in the flat region. The coalescence of the jet takes a longer time than in the baseline case but results in a faster moving jet in the steady state.

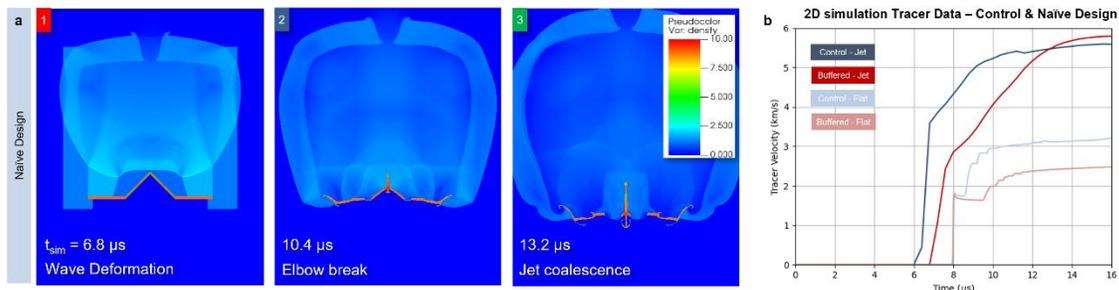

Figure 4: (a) Images 1, 2 and 3 show the time evolution of the naïve buffered charge simulation in density. (b) Tracer velocity results for the naïve charge are shown in red and compared to the baseline results in blue.

### 3.1.1. INVERSE DESIGN RESULTS

The best performing design found by the computational optimization algorithm produces a buffer as can be seen in Figure 2d. Simulations results for the optimized design are shown in Figure 5. This design yields an objective function value of 6.52, composed of a maximum jet velocity of 3.51 km/s and a maximum flat velocity of 2.57 km/s. For reference, the objective function value of the baseline case is 7.30 with a jet velocity of 5.60 km/s and flat velocity of 3.23 km/s. Thus, we see a 37% reduction in jet velocity with only a 20% reduction in flat velocity corresponding to a 26% reduction in HE mass. One can see that the buffer not only reduces the jet velocity significantly more than the change in HE mass, but the buffer also enhances the flat velocity such that the flat velocity reduction is less than the HE mass reduction.

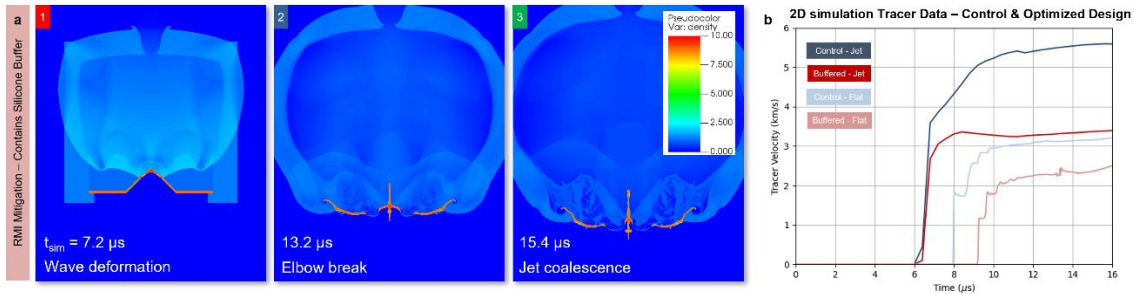

Figure 5: (a) Images 1, 2 and 3 show the time evolution of the computationally optimized buffered charge simulation in density. (b) Tracer velocity results for the optimized design are shown in red and compared to the baseline results in blue.

At first glance, the visual appearance of the buffer is counter intuitive. Immediately behind the jetting region of the liner, there is only a very thin layer of inert, with the bulk of the buffer lying behind the region joining the flat and 'V' regions of the liner. There is also a small region of the buffer where it thins behind the flat of the liner, giving an inlet for the HE to fill. It is believed that this inlet enhances the push on the flat of the liner where the velocity is measured. The mechanisms behind the jet velocity reduction, however, are more complex. One can determine from the tracer velocity plot that the initial rise in jet velocity due to the primary interaction with the apex of the liner and the detonation front is nearly the same with and without the buffer, as expected from local liner thickness. The jet velocity is modified after the initial rise due to a confluence of factors including the wavefront deformation caused by the buffer geometry, the change in the transfer of kinetic energy from the detonation through the buffer into the liner, and the buffer's action as a vorticity sink. First, the wavefront deformation within the buffer serves to change the angle at which the kinetic energy from the detonation wave interacts with the 'V' section of the liner. We propose that this particular buffer geometry changes the interaction angle in a way that discourages the coalescence of the jet through the change in horizontal energy deposit into the liner. In addition to the interaction angle being changed, the buffer absorbs the kinetic energy from the detonation in such a way that we see large regions of lower kinetic energy, as compared to the detonation front, within the buffer following the interaction of the detonation front with the buffer's boundary. While we saw in the jet augmenting case above that this region of kinetic energy results in a more sustained energy transfer into the liner, the geometry of the buffer here results in a more sustained energy transfer into the flat region of the liner with a weaker effect occurring behind the 'V' of the liner. Due to the thickness of the buffer behind the 'V' of the liner, the energy dissipation of the buffer is more significant than the sustain of the energy transfer there. These two mechanisms' effects on the jet velocity evolution are most pronounced following the first kink in the velocity plot. We see a weaker velocity increase due to the change in the kinetic energy transfer into the liner. As the 'V' section of the liner coalesces into the jet, we see an initial velocity increase followed by a velocity decrease which then evolves into a quasi-steady state. It is the jet coalescence phenomenon that controls this mid- and late-time behavior of the jet. We propose that the jet coalescence is modified by the buffer acting as a vorticity sink. As can be seen in images 2 and 3 of the time evolution of the jet in Figure 5, the buffer immediately behind the 'V' region of the liner retains a relatively high density throughout the simulation. A closer look at the baroclinic term of the vorticity equation within the device, Figure 6, shows significantly smaller regions of high vorticity in the liner when the buffer is included in the device. Due to the smaller regions of vorticity, the rays of the 'V' section of the liner are pulled in significantly slower when the buffer is included. The reduction in size of the high vorticity regions is likely caused by the deformation of the buffer capturing the energy of the vortex that would otherwise be used to deform the liner. This is in contrast to the mechanism found in Sterbentz et. al [11] where they used a buffer mechanism to intentionally create an upstream interface instability which counteracted the growth of RMI.

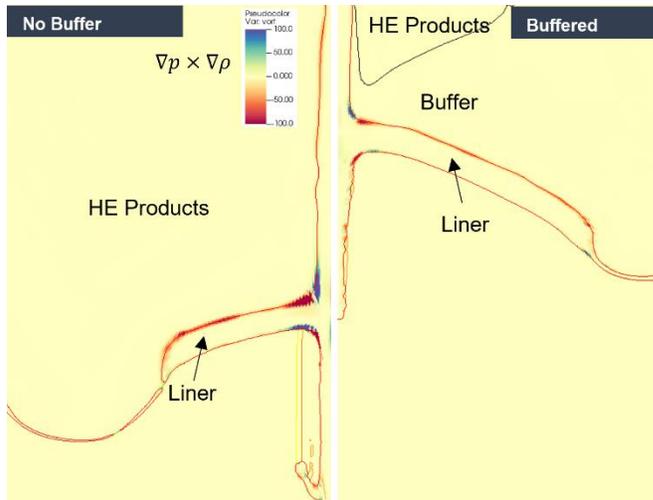

Figure 6: A comparison plot of the baroclinic term of the vorticity equation for the charge with no buffer (left) and the charge with machine-learned buffer (right).

### 3.1.2. 3D SIMULATION RESULTS

While we utilized 2D simulations to generate the buffer geometry and understand the physics of the jet evolution, we also performed 3D simulations to better capture the experimental design and to be able to capture the parallax effect in the flash X-ray images. Results of these simulations may be found in Figure 7 and Figure 8, along with discussions of the results there. Although the resolution of these simulations is significantly lower, the experimental radiographs are captured with good accuracy for both experiments. While there are significant differences between the 2D and 3D simulations, especially in the trace velocities, the agreement between the 3D simulations and the experimental results is very good. As the only significant difference between the simulations in 2D and 3D is the inclusion of edge effects, we must believe that these effects are dramatic enough to explain the differences between simulations.

## 3.2. DETONATION EXPERIMENTS

### 3.2.1. PART FABRICATION AND INSPECTION

As discussed in previous sections, both components in the experiment were prepared using an additive manufacturing process. The manufacturing process for the baseline experiment proceeded as expected with only minor changes in flowrates made throughout the print to ensure that the correct amount of material was being dispensed. However, during the fabrication of the RMI-mitigating experiment, a small defect was introduced when the nozzle of the silicone dispenser was obstructed. This resulted in material misplacement in the top 1 mm section of the silicone component and mixing of silicone in the HE layer printed immediately after (Z=18-20 mm). The HE mass of the final components was 209 g and 154 g for the baseline and RMI mitigation experiments, respectively. It is important to note that the HE mass in the RMI mitigation experiment is ~26% less than the baseline since much of the volume was replaced by the inert buffer, however, the final expected mass for both experiments closely matched the predicted mass given the density of the materials used. The impact of the reduced HE mass on the expected results will be discussed in a later section.

After curing, the components were inspected for voids and major defects using XRCT. A detailed analysis on the part quality below Z=10 mm is particularly difficult due to beam hardening of the X-ray in the copper liner and the challenges that arise from this in digital reconstruction. The baseline experiment did not exhibit any noticeable voids in the final part after curing, suggesting that the material was adequately degassed and appropriately filled the mold during the print. However, the RMI mitigation experiment did have voids and defects present in the part. As discussed above, a defect was introduced to the part during the fabrication process due to an obstruction. It was determined that this defect would have a negligible impact on the experimental results, though, as the raw radiographs still closely resembled the intended part. Further inspection of the digital reconstruction revealed multiple air bubbles in the layers containing both silicone and HE (~Z=10-19) and a large void near the edge face of the component spanning ~2 mm in the Z direction. It is believed that the silicone, which had a significant amount of trapped air, was not fully degassed prior to printing. This gas later migrated and expanded in the elevated temperatures the component was exposed to in the curing process and subsequently collected to form a larger void. It was ultimately determined that these voids would also have a relatively small impact on the performance of the part, especially considering their placement towards part's exterior, although a formal analysis is reserved for future studies. Another minor defect spanning ~Z=9.5-10 mm can be seen as the HE protrudes into the silicone layer below it. This is likely due to the weight of the HE and the pressure of the material above it since the HE is much denser than the silicone (~1.7 g/cm3 vs. ~1.1 g/cm3). This minor defect was also considered to have a negligible impact on the performance of the part in the region of interest.

### 3.2.2. FLASH X-RAY IMAGES

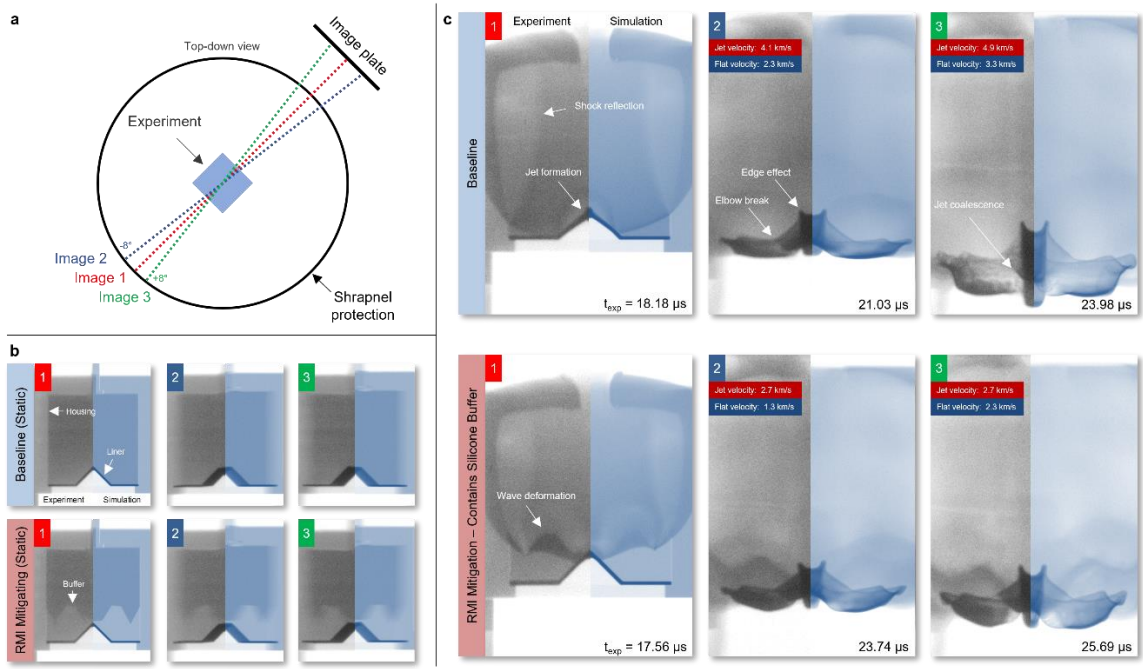

Figure 7: Flash X-ray experiment configuration and results. (a) Top-down view of the shrapnel catcher for the experiment. Note that the drawing is not to scale and the X-ray collimators are not shown. (b) Static X-ray images for the baseline experiment (top) and RMI mitigation experiment (bottom). (c) Flash X-ray images captured during baseline detonation experiment (top) and RMI mitigation experiment (bottom). Numbers in the upper right corner of the images correspond to the head/image number in a. Gray image is the raw radiograph and blue overlay is a simulated radiograph generated with 3D hydrodynamic simulation results.

Flash X-ray (FXR) images were the primary diagnostic used to evaluate instability mitigation and the effect of the silicone buffer on the detonation event. As discussed in the previous sections, a linear shaped charge design was chosen to better capture the early time material deformation dynamics, however the experimental setup would lead to a parallax effect for images taken off-center.

Prior to the detonation experiment, static radiographs of the assembly were taken on the shot stand to evaluate image quality and evaluate the parallax effect for radiographs taken off the center axis. Figure 7b shows the static radiographs for the baseline (top) and RMI mitigation experiments (bottom). The image down the centerline of the parts (Image 1) had clearly-resolved assembly features, especially including the placement of the silicone buffer in the RMI mitigating configuration. Images 2 and 3 of each assembly highlight the impact of the parallax on the radiographs. Features in these components became much more difficult to resolve, however, these images provide an oblique viewing angle that would prove effective for capturing edge effects in the detonation experiment. To better compare captured images to predictions, simulated radiographs (overlays in Figure 7) for the static images were generated in HADES using models of the individual parts, material isotope fraction, X-ray emission spectrum, and shot assembly layout. The simulated radiographs closely matched the captured static images with minor differences largely arising from differences in resolution and noise artifacts introduced from brightening the radiographs.

Three radiographs would be captured for each experiment and the timing for the FXR pulses was determined based on key features of the detonation event from the 2D simulations. For the baseline experiment, timing was chosen to capture (1) the initial breakout of the jet at the HE/liner interface, (2) the break at the "elbow" of the liner that would separate the jet from the flat region, and (3) the final jet coalescence. Timing for the RMI mitigation experiment was slightly altered to capture (1) evidence of shock wave deformation shortly after passing the HE/silicone buffer interface, (2) the "elbow" break in the liner, and (3) final jet coalescence. Images 2 and 3 would also be used to estimate jet velocity given the travel distance of the jet and the time between the images being captured.

Radiographs from the baseline experiment can be seen in Figure 7c in the top image row. Image 1 of the baseline experiment clearly shows the initial jet formation of the experiment, as well as a reflected shock wave from the wall and liner back into the gaseous products. Image 2 also captured the features of interest, notably including the evolution of the jet and the elbow break. Due to the parallax effect of the image being taken from an oblique angle, the "elbow break" appears as a lighter region on the deformed liner when projected. Image 2 also shows rounded features in the jet and slug of the liner which stem from edge effects in the detonation event. These edge effects are the result of energy loss into the case and surrounding area and cause a lagging edge in the detonation front. This lagging edge, in turn, also leads to the appearance of a curved jet front when projected into 2D radiograph. Image 3 of the baseline experiment also captured the jet displacement, and we believe that the final jet coalescence appears as a smeared bulb of material protruding approximately halfway down the jet length. Velocity of the jet and flat were estimated using ImageJ [24] whereby the distance from the PDV plate to the tip of the jet and lowest portion was measured for each image, the displacement from the previous image was calculated, and the time between frames was given by the FXR trigger feedback. The jet velocity was estimated to be 4.1 km/s at the time image 2 was captured and further increased to 4.9 km/s at the capture time of image 3. Similarly, the flat region of the baseline experiment accelerated from 2.3 km/s in image 2 to 3.3 km/s in image 3.

Again, to compare the experimental results to predictions, 3D hydrodynamic simulation data was imported and simulated radiographs for the experimental setup were generated using HADES. Simulated radiographs are shown as blue-colored overlays in the baseline FXR images. The simulated radiographs for the experiment are nearly identical to the captured radiographs and show the same key features of the detonation event. Deviations between the predictions and experiment are largely due to slight differences in the predicted detonation event timing. This is particularly noticeable with the absence of a jet in the simulated radiograph, however, the other images taken at later times more closely resemble the experiment.

The radiographs from the RMI mitigating design experiment, shown in the bottom image row of Figure 7c, significantly differed from the baseline experiment. More importantly, though, these radiographs demonstrate the effectiveness of the silicone buffer as an instability mitigation technique. Image 1 of the second experiment captured the intended shock wave deformation, but does not show early breakout of the jet in the experiment. Image 2 depicts the jet evolution and evidence of the jet coalescence can be seen towards the edge of the jets. Similar to the baseline experiment, image 3 also has a smeared bulb towards the midway point of the jet where the material coalesces. The jet velocity of the RMI mitigating design was estimated to be 2.7 km/s in both images 2 and 3, however, the velocity of the flat region increased from 1.3 km/s in image 2 to 2.3 km/s in image 3. Simulated radiographs for the RMI mitigating design closely match the experimental results, further confirming that the component performed as expected.

There are multiple key differences between the two experiments' radiographs and the velocity estimations that capture the impact of the silicone buffer on RMI mitigation. Radiographs of the two experiments exhibit very different material deformation and jet evolution behaviors. The flat regions of the RMI mitigating design appear more curved in both images 2 and 3 compared to the baseline case. The jet also appears much farther ahead of the flat region in the last image of the baseline experiment whereas the jet appears in-line with the tips of the flat region in the RMI mitigating design. The peak jet velocity in the RMI mitigating design was ~47% slower than the baseline experiment. The reduction in jet velocity is much larger than the reduction in the HE mass between the two designs (~24%). The flat region of the experiment, which was included as a reference region for velocity, scaled more closely with the HE mass (~30% velocity reduction). The relatively large change in the jet velocity compared to the change in the HE mass and flat velocity serves as further confirmation that the jet reduction was a direct effect of the ML-designed silicone buffer optimized to reduce RMI.

### 3.2.3. PHOTONIC DOPPLER VELOCITMETRY

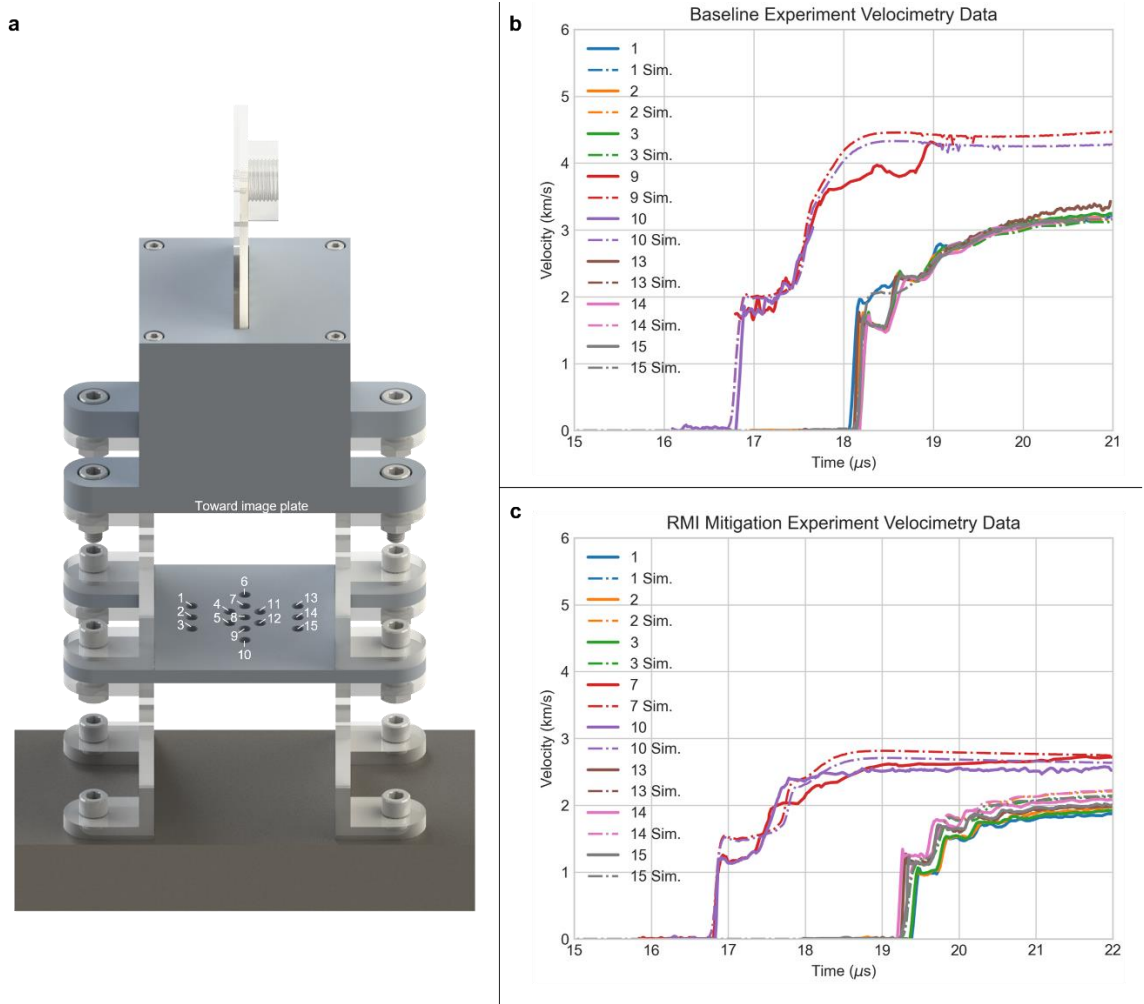

Figure 8: Photonic doppler velocimetry data capture and results. (a) Probe placement and number in experiment assembly. (b) Baseline experiment velocimetry data. (c) RMI mitigation experiment velocimetry data. Please note that not all probe data is shown due to either poor signal-to-noise ratios or loss of data (described in text).

In addition to the images captured by FXR, photonic doppler velocimetry (PDV) was used as a quantitative tool to measure material deformation and velocities with higher time resolution. These results would also be used to corroborate the results observed in the FXR images. The layout of the probes in the experiment can be seen in Figure 8a. Collimating fiber optic probes were used to capture velocimetry data over a larger spatial and temporal domain than could be captured using fixed focal length probes. Multiple probes were also placed along the same axis in the part for redundancy as the material deformation at later times can lead to loss of signal. Redundancy in probe placement is also useful if data collected from similar locations has a poor signal-to-noise ratio (SNR) or probe malfunctions. Probes 1-3 and 13-15 were placed along the flat face of the liner and probes 6-10 were placed along the centerline of the liner where the jet would form. Probes 4, 5, 11, and 12 were not evaluated in this study since the data collected could not be readily compared to tracer data predicted in hydrodynamic simulations.

Unfortunately, while integrating the probes into the shot assembly, it was discovered that the signal-to-noise ratio for the reflected light in many of the probes was considerably low. It was later determined that this was caused by the bead-blasted surface finish of the copper liners; the surface finish scattered a significant amount of light and consequently reduced the amount of light being reflected into the PDV probes. Future iterations of this experiment will use liners with a more reflective surface finish to improve the signal return. It is important to note that conducting dynamic experiments with PDV is inherently difficult due to the tradeoffs between surface reflectivity and specularity. A highly reflective surface will have high SNR if the accelerated material is moving perfectly normal to the probe and the orientation is maintained, however a specular surface finish will have a lower SNR but will provide some signal if not perfectly aligned to the probe. We chose to pursue a more specular surface finish to maximize the chance of getting a signal during the detonation test.

Velocimetry data for the control experiment can be seen in Figure 8b along with the PDV results predicted from 3D hydrodynamic simulations. An early jump in velocity from 0 to 1.8 km/s began at ~16.8 µs in the probes capturing data along the jetting region of the part. Shortly thereafter, there is a plateau and second jump in the jet velocity. The peak jet velocity recorded in the control experiment was 4.3 km/s. It is, however, important to note that the quality of the data recorded along the centerline of the part was relatively poor with multiple probes recording either very limited or no data. Conversely, PDV data collected along the flat regions of the components had considerably better signal returns compared to data in the jet (likely because the jet broke apart or was diverted out of the laser beam path). The velocity jump in the flat region began 18.1 µs (delayed 1.3 µs from the jet because of the additional distance traversed) and quickly rose to ~1.7-2.0 km/s by 18.2 µs. After the shock reflections subsided, the flat region probes recorded a terminal velocity of 3.2-3.4 km/s. This data closely matches the 3D hydrodynamic simulation-predicted velocities. Deviations in the data from probe-to-probe is likely due to edge effects in the detonation event, which is difficult to accurately predict in 3D simulations. The velocimetry data also matches the velocity estimated from the radiographs at later times, however, it is difficult to reliably compare the jet velocimetry data to the radiographs given the poor signal.

The velocimetry data recorded for the RMI mitigation experiment had much better signal-to-noise ratios and yielded more reliable results over longer time periods (see Figure 8c). Like the control case, the initial jump off in the jet velocity began at 16.8 µs, however first peak in the jet velocity trace only rose to 1.2 km/s. This is notably lower than the initial velocity peak predicted by the hydrodynamic simulations (1.55 km/s), although the deviation could be due to edge effects or the slight defects introduced during the print. The highest recorded jet velocity until $t_{exp}$ = 22 µs was 2.8 km/s, although low-quality data captured at later times peaked at 3.0 km/s. Velocity jumps in the flat regions began at 19.2 µs, which was delayed 2.4 µs from the jump in the jet velocity. The delay in the RMI mitigation experiment was longer than that in the control experiment since the shock propagation velocity is significantly reduced by the silicone insert and the reduced amount of explosive driving the shock. The velocity of the flat region in the RMI mitigation experiment was 1.9-2.1 km/s, which is in relative agreement with the predicted velocities of 2.2 km/s. Both the jet and flat region velocities measured via PDV match the terminal velocities estimated from the radiographs.

The reduction in jet velocities between these experiments further confirms that the ML-designed silicone buffer was effective in reducing the RM instability. The jet velocity reduction was 37% at early times in the experiment whereas the HE content was only reduced by 24%. However, given the poor signal-to-noise ratios, the comparison of velocities over the duration of the experiment should be relegated to a combination of simulation data and radiography results. Future iterations of this experiment will use liners with a more reflective surface finish to improve the signal return and yield high quality data over longer experiment times.

## 4. CONCLUSIONS

In this work, we demonstrated control of the Richtmyer-Meshkov instability in a linear shaped charge analogue by using inverse design to reduce the jet velocity. A genetic algorithm was used on a parameterized hydrodynamic simulation to optimize the shape of a silicone buffer in an explosive charge to reduce jet velocity in an arbitrarily designed linear shaped charge. After optimization of the silicone buffer, explosive components were prepared via a multi-material additive manufacturing process. Detonation experiments of these components showed that the jet velocity was reduced as confirmed by both flash X-ray radiography and photon doppler velocimetry. The reduction in jet velocity was also greater than the reduction of the energy in the system, further confirming that the shape of the silicone primarily influenced the observed effects on jet velocity. This study shows that counterintuitive results generated by inverse design can be realized with advanced manufacturing techniques and may be experimentally valid, suggesting that this approach can be applied to more complex physics problems.

## DECLARATION OF COMPETING INTERESTS

The authors declare that they have no known competing financial interests or personal relationships that could have appeared to influence the work reported in this paper.

## AUTHOR CONTRIBUTIONS

D.J. Kline: conceptualization, methodology, software, resources, formal analysis, investigation, resources, data curation, writing – original draft, visualization, supervision, project administration. M.P. Hennessey: conceptualization, methodology, software, resources, formal analysis, investigation, data curation, writing – original draft, visualization. D.K. Amondson: conceptualization, methodology, investigation, resources, data curation, writing – review & editing. S. Lin: resources, investigation. M.D. Grapes: software, resources, writing – review & editing. M. Ferrucci: software. P. Li: software. H.K. Springer: methodology, software, validation, writing – review & editing, supervision. R.V. Reeves: methodology, validation, supervision. K.T. Sullivan: writing – review & editing, supervision, project administration, funding acquisition. J.L. Belof: conceptualization, methodology, resources, writing – review & editing, supervision, project administration, funding acquisition.

## ACKNOWLEDGMENTS

We would like to acknowledge the significant amount of work performed by the High Explosives Application Facility staff in support of this work, especially including Anthony B. Olson, Jason G. Lumanlan, Joel Alfonso, Caleb A. Glickman, Ferdinand Dizon, Deanna M. Kahmke, James M. Clark, and Doug Lahowe. This work was performed under the auspices of the U.S. Department of Energy by Lawrence Livermore National Laboratory (LLNL) under Contract DE-AC52-07NA27344. We gratefully acknowledge the LLNL Lab Directed Research and Development Program for funding support of this research under Project No. 21-SI-006. Document release number LLNL-JRNL-855667.